\documentclass[authoryear,preprint,3p,12pt]{elsarticle}

\usepackage{amssymb}

 \usepackage{lineno}

\journal{Planetary and Space Science}

\begin{document}

\begin{frontmatter}


 \title{SPH calculations of asteroid disruptions: The role of pressure dependent failure models}


\author[label1]{Martin Jutzi}
\address[label1]{University of Bern,
   Center for Space and Habitability,
   Physics Institute,
   Sidlerstrasse 5,
     3012 Bern, Switzerland}
\ead{martin.jutzi@space.unibe.ch}




\begin{abstract}

We present recent improvements of the modeling of the disruption of strength dominated bodies using the Smooth Particle Hydrodynamics (SPH) technique. The improvements include an updated strength model and a friction model, which are successfully tested by a comparison with laboratory experiments. In the modeling of catastrophic disruptions of asteroids, a comparison between old and new strength models shows no significant deviation in the case of targets which are initially non-porous, fully intact and have a homogeneous structure (such as the targets used in the study by \citealt{BA1999}).
However, for many cases (e.g. initially partly or fully damaged targets, rubble-pile structures, etc.) we find that it is crucial that friction is taken into account and the material has a pressure dependent shear strength. Our investigations of the catastrophic disruption threshold  $Q^*_{D}$ as a function of target properties and target sizes up to a few 100 km show that a fully damaged target modeled without friction has a $Q^*_{D}$ which is significantly (5-10 times) smaller than in the case where friction is included. When the effect of the energy dissipation due to compaction (pore crushing) is taken into account as well, the targets become even stronger ($Q^*_{D}$ is increased by a factor of 2-3). On the other hand, cohesion is found to have an negligible effect at large scales and is only important at scales $\lesssim$ 1km. 

Our results show the relative effects of strength, friction and porosity on the outcome of collisions among small ($\lesssim$  1000 km) bodies. These results will be used in a future study to improve existing scaling laws for the outcome of collisions (e.g. \citealt{LS2012}). 
\end{abstract}


\end{frontmatter}


\section{Introduction}\label{s:intro}

Collisions play a fundamental role in the formation and evolution of the planets and the small body populations of the Solar System. Models of the  evolution of such populations (e.g. the Asteroid Belt) compute the time dependent size and velocity distributions of the objects as a result of both collisional and dynamical processes. A scaling parameter often used in such numerical models is the critical specific impact energy $Q^*_D$, which results in the escape of half of the target's mass in a collision. The parameter $Q^*_D$ is called  the catastrophic impact energy threshold (also called the dispersion threshold). The specific impact energy is often defined as $Q = 0.5 m_p v_p^2 / M_T$, where $m_p$, $v_p$ and $M_T$ are the mass and speed of the projectile and the mass of the target, respectively. The catastrophic disruption threshold $Q^*_D$ is then given by the specific impact energy leading to a largest (reaccumulated) fragment $M_{lr}$ containing 50\% of the original targetÕs mass. In recent studies (e.g. \citealt{SL2009,LS2012}), a more general definition of the specific impact energy was proposed which also takes the mass of the impactor into account:
\begin{equation}
Q_R=\frac{0.5 m_p v_p^2+0.5 M_T V_T^2}{M_{tot}}=\frac{0.5 \mu V_i^2}{M_{tot}}
\end{equation}
where $M_{tot}=m_p+M_T$ and $\mu_i=m_p M_T / M_{tot}$ and $V_i$ is the relative velocity. The corresponding radius $R_{C1}$ is defined as the spherical radius of the combined projectile and target masses at a density of 1 g cm$^{-3}$. According to this new definition, the catastrophic disruption threshold is then called $Q^*_{RD}$. 

Values of $Q^*_D$ (or $Q^*_{RD}$) have been estimated using both laboratory and numerical hydrocode experiments (see e.g. \citealt{Ho2002,As2002}).
For the small body populations, the first suite of numerical calculations aimed at characterizing the catastrophic disruption threshold in both the strength regime and the gravity regime was performed by \citet{BA1999}, who used a  smoothed particle hydrodynamics (SPH) code \citep{BA1994,BA1995} to simulate the breakup of basalt and icy bodies from centimeters-scale to hundreds kilometers in diameter. More recently, \citet{LS2009}  computed $Q^*_D$ curves using the hydro code CTH \citep{Mc1990} to compute the fragmentation phase and the N-body code \emph{pkdgrav} \citep{Ri2000} to compute the subsequent gravitational evolution of the fragments. In this study, the dependency of $Q^*_D$ on the strength of the target was investigated. In a recent study by \citet{Ju2010},  the effect of target porosity on $Q^*_D$ was investigated using an extended version of the SPH code \citep{Ju2008}.  In this study, the size and velocity distribution of the fragments was computed as well, using the \emph{pkdgrav} code. \citet{Be2012} performed a study of a large number of  collisions among $R_t$ = 50 km rubble pile bodies using the original SPH code by  \citet{BA1994,BA1995}. 

A numerical tool which is very suitable and has been often used to study disruptive collisions among rocky bodies in general, and was used in many of the above-mentioned asteroid disruption studies, is based on the SPH method. Over the last decades, the basic method has been extended by implementing additional physics (e.g.  \citealt{BA1994,BA1995,Ju2008,Ju2009}) with the goal to realistically model rocky bodies with various internal structures. In addition to improved constitutive models,  models which mimic the complex macroscopic structure of rubble pile-like bodies have been used as well (e.g. \citealt{As1998,Be2012}). 
Although the SPH models used in planetary sciences have been significantly improved over the last decades, they were still lacking aspects that can be important in some impact regimes. For example, previous SPH strength models used the so-called von Mises yield criterion, which does not describe well the behavior of rocky materials which are known to have a pressure dependent shear strength. Furthermore, fully damaged material was treated as a strengthless 'fluid' in previous models. Finally, while self-gravity is included in the versions of the SPH codes used to model giant collisions, it is often not implemented in the SPH code versions which also include the physics of solid bodies. 

In this paper, we present improvements of the SPH technique concerning the modeling of the disruption of strength dominated bodies. These improvements include a pressure dependent Drucker-Prager-like yield criterion, and a friction model for the damaged material. A few test cases are presented. Using the improved models, we then systematically study the effects various target properties (strength, porosity and friction) on the outcome of a disruptive collision  ($Q^*_D$). In this study, we use targets with a homogeneous internal structure (i.e., it is assumed that the voids or inhomogeneities are sufficiently small that their distribution can be assumed uniform and isotropic over the relevant scales). 
In section 2, we present our numerical tool and the recent improvements and show two test cases. In section 3, the catastrophic disruption threshold $Q^*_D$  is investigated as a function of material properties. In section 4, the results are discussed and future work is indicated.

\section{Modeling} \label{s:method}
\subsection{Previous SPH models}
\citet{BA1994,BA1995} extended the standard gas dynamics SPH approach to include an elastic-perfectly plastic material description (see, e.g. \citealt{LP1991}) and a model of brittle failure based on the one of \citet{GK1980}. In the fracture model, a state variable D (for damage) was introduced which expresses the reduction in strength under tensile loading and which varies between $D$ = 0 and $D$ = 1. Damage accumulates when the local tensile strain reached the activation threshold of a flaw. As stress limiter, the von Mises yield criterion was used. Finally the so-called Tillotson equation of state for basalt \citep{Ti1962} was used to relate the pressure to the density and the internal energy.  We refer the reader to the papers by \citet{BA1994,BA1995} for a detailed description of this method. This code was then used  for instance by \citet{BA1999} to make a first complete characterization of $Q^*_D$  for basalt and ice targets at different impact speeds. The same version (in terms of material models) of the SPH code  was used by \citet{Be2012} to study collisions among rubble pile asteroids. 
\citet{Ju2008} extended the \citet{BA1994,BA1995} method by implementing a sub-resolution porosity model based on the P-alpha model \citep{He1969,CH1972}. In this implementation, a distention parameter $\alpha=\rho_s/\rho$ is introduced where $\rho$ is the bulk density of the porous material and $\rho_s$ is the density of the corresponding solid (matrix) material. The distention $\alpha$ is defined as a function of pressure via the so-called crush-curve.  It is used in the EOS to compute the pressure as a function of the matrix density. The porosity model was successfully tested by a comparison to laboratory experiments involving porous pumice \citep{Ju2009}. The SPH code used by \citet{Ju2008} and in the later studies is a parallelized version \citep{Ny2004} of the original code by \citet{BA1994,BA1995}.

\subsection{Recent improvements}\label{sec:improv}
In the original implementation of the strength and fracture model by \citet{BA1994,BA1995}, the pressure independent von Mises yield criterion was used. However, it is known that the shear strength of rocks is pressure depended (i.e., it increases with increasing confining pressure). A pressure dependent yield criterion often used to deal with rocky materials is the  Drucker - Prager  yield criterion

\begin{equation}
\sqrt{J_2}+\alpha_\phi I_1-k_c = 0
\end{equation}
where $I_1$ is the first invariant of the stress tensor, $J_2$ is the second invariant of the deviatoric stress tensor $\alpha_\phi$, and $k_c$ are Drucker Prager constants, which are related to the Coulomb's material constants (coefficient of friction $\mu$ and cohesion $Y_0$)  (see e.g. \citealt{Bu2008}). For the implementation in our SPH code, we use $\sqrt{J_2}$ as a measure of the stress state, and we define a pressure dependent yield strength $Y_i$ for intact rock following \cite{Co2004}:
\begin{equation}
Y_i = Y_0+ \frac{\mu_i P}{1+\mu_i P/(Y_M-Y_0)}
\label{eq:Yi}
\end{equation}
where $Y_0$ is the shear strength at $P$ = 0 and $Y_M$ is the shear strength at $P$ = $\infty$ and $\mu_i$ the coefficient of internal friction. 
 As in the previous model, $Y_i$ is temperature depended:
\begin{equation}
Y_i \rightarrow Y_i\left(1-\frac{u}{u_{melt}}\right)
\end{equation}
where $u$ is the specific internal energy and $u_{melt}$ the specific melting energy.

In the original model by \citet{BA1994,BA1995}, fully damaged material was treated as strength-less ("fluid"). As we shall see in section \ref{cliffcol}, this simplification leads to reasonably accurate results in the case of disruptive collisions between initially intact bodies. However, in may situations it is important to take into account the friction in the modeling of fully damaged (i.e, granular) material. This is certainly the case when we want to study collisions between rubble pile like, granular bodies, or to study the finally shape of a body after an impact, and it is also expected to be important in the cratering regime of impacts. 
To model  fully damaged rock (damage $D=1$), which includes granular material in general, we use a yield strength
\begin{equation}
Y_d = \mu_d P
\label{eq:Yd}
\end{equation}
where $\mu_d$ is the coefficient of friction of the damaged material \citep{Co2004}. Note that  $Y_d$ is limited to $Y_d\le Y_i$. In case the modeling starts with a intact or partially damaged material, a smooth transition between the criterions (\ref{eq:Yi}) and (\ref{eq:Yd})  is used:
\begin{equation}
Y=(D-1)Y_i + D Y_d
\label{eq:combined}
\end{equation}
where $Y$ is limited to $Y\le Y_i$.

If the measure of the stress state $\sqrt{J_2}$ exceeds $Y$, the components of the deviatoric stress tensor are reduced by a factor $Y/\sqrt J_2$. The yield strength given by (\ref{eq:combined}) replaces the von Mises yield criterion used in the previous SPH models. 

Finally, to compute the accumulation of damage $D$,  a tensile brittle failure model based on the one of Grady and Kipp (1980) is used (see \citet{BA1994,BA1995,Ju2008}. Damage accumulates when the local tensile strain $\epsilon_i$ reached the activation threshold of a flaw. Note that $\epsilon_i$ is obtained from the maximum tensile stress $\sigma^t_i$ after a principal axis transformation.

The friction model described above (equation \ref{eq:Yd}) assumes a constant friction coefficient. However,  \citet{Jo2006} use a model with a rate dependent friction coefficient to reproduce experiments of dense granular flow. Their 3D model is based on a Drucker-Prager like yield criterion and a friction coefficient which is a function of the inertial number, defined as
\begin{equation}
I = |\dot\gamma| d/ (P/\rho_s)^{0.5}
\label{eq:I}
\end{equation}
where $P$ is the isotropic pressure,  $|\dot\gamma|$ the second invariant of the strain rate tensor $\dot\gamma_{ij}$, $d$ is the particle size and $\rho_s$  is the particle density. Using $I$, the following law for a strain rate dependent friction coefficient was proposed
\begin{equation}
\mu(I)=\mu_s+(\mu_2-\mu_s)/(I_0/I+1)
\label{eq:muI}
\end{equation}
where $\mu_s$ is the critical value of the friction coefficient at zero shear rate, $\mu_2$ is the limiting value at high $I$, and $I_0$ is a constant. 
We implemented the relation (\ref{eq:muI}) in the SPH code by replacing the constant coefficient $\mu_d$ in equation (\ref{eq:Yd})  by $\mu(I)$. However, as we shall see in section \ref{cliffcol}, it is not clear whether or not a strain rate depend friction coefficient allows to better reproduce the behavior of dense granular material. For this reason, our model uses a constant value 
\begin{equation}
\mu(I)=\mu_d
\label{eq:muIconst}
\end{equation}
 unless indicated otherwise. 

\subsection{Tests}
In this section, we present two test cases where we compare our model with experimental results. In the first case, we test the ability of our numerical tool to deal with fully fragmented (i.e., granular) material. For this, the collapse of a cliff of granular material is simulated. In the second case, we test the updated strength models by a comparison to laboratory impact experiments using porous gypsum targets \citep{Ok2009}.
\subsubsection{Cliff-collapse}\label{cliffcol}
We model the collapse of a cliff of granular material and we use the results of experiments of \citet{La2005} to verify our friction model. Note that \citet{Ho2013} performed a detailed study of the same problem using a continuum CTH code and Mohr-Coulomb and/or Drucker-Prager models. 
In the cliff collapse problem (see \citealt{Ho2013} for a detailed description), a granular material is initially constrained in a rectangular region with height $H_0$ and length $L_0$. The width is assumed to be large and assumed to be of no consequence. At the time $t=0$, the wall on one side is removed and the material begins to flow down due to downward gravity $g$. Eventually, the flow ceases in a final run-out configuration with a with maximum height H and length L which depend on the properties (angle of friction) of the granular material. As it was found experimentally (e.g. \citealt{La2005}) as well as by a scaling analyses \citep{Ho2013}, the final scaled profile only depends on the initial height to length ratio and the angle of friction. The gravity and the height of the cliff only affect the time and length scales of the problem. This allows to model the problem using artificially large dimensions to avoid problems related to time-step restrictions (see Holsapple 2013). In the 3D calculation presented below, the dimensions (including the particle size $d$) are increased by a factor $\sim 10^5$.

\begin{figure}
\begin{center}
\includegraphics[width=14cm]{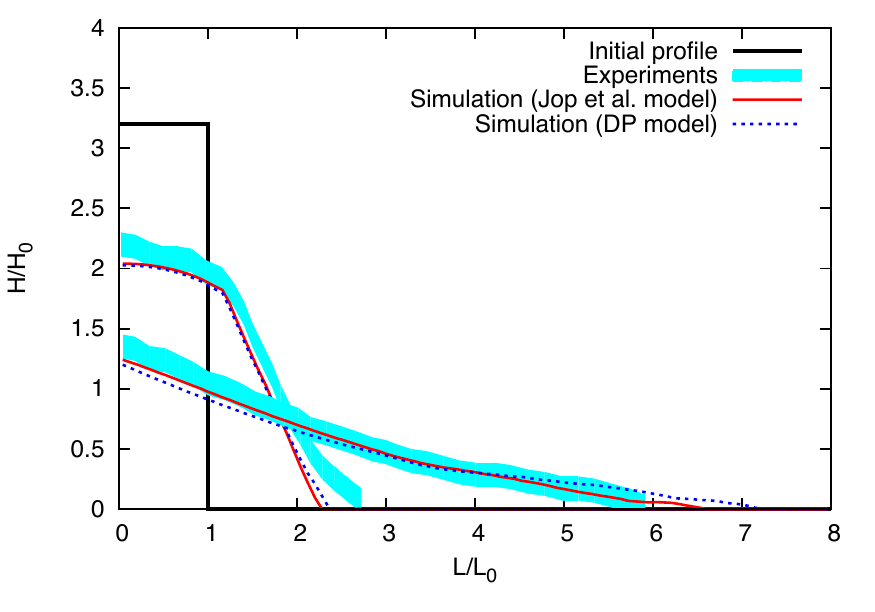}
\caption{Collapse of a granular cliff. The height (y-axis) is normalized by the initial height $H_0$ and the length (x-axis) is normalized by the initial length $L_0$. The experimental results by \citet{La2005} (light blue) are compared to the results of SPH code calculations using the model for dense granular flow by \citet{Jo2006} (solid red line) and a Drucker-Prager friction model with a constant friction coefficient (dotted blue line). The profiles are shown at the times $t = \tau = \sqrt{ H_0/g}$ and $t\to\infty$ (final profiles).
}\label{fig:profiles}
\end{center}

\end{figure}

Figure \ref{fig:profiles}  shows a SPH code calculation of a cliff collapse compared with the results of the experiments by \citet{La2005}. In the experiments, glass beads of different sizes, and various initial height to length ratios were used. When scaled by the characteristic length $L_0$ and time $\tau= \sqrt{H_0/g}$, it was found that the profile curves of the different experiments are barley distinguishable. These measured curves are represented by the blue line in Figure \ref{fig:profiles} at $t = \tau = \sqrt{ H_0/g}$ and $t\to\infty$ (final profiles). 

The results of our simulation using two different laws for $\mu(I)$ (Equations \ref{eq:muI} and  \ref{eq:muIconst}) are also shown in Figure \ref{fig:profiles}. The parameters used in the first case are $\mu_s$ = tan(20.9), $\mu_2$ = tan(32.76) and $I_0$ = 0.279  \citep{Jo2006}, and we use $\mu_d=\mu_s$ in the second case (constant friction coefficient). As it can be seen, both models reproduce very well the experimental results and there is no large difference between the model with a constant $\mu_d$ and a rate dependent $\mu(I)$. The reason is that in the cliff collapse problem studied here, the inertial number stays small and therefore $\mu(I)\sim \mu_s=\mu_d$. Our results suggest that in the flow regime investigated by this kind of experiment, both models work equally well, which also means that the global outcome of such events is well reproduced by using a single parameter $\mu_d$.

\subsubsection{Impact experiment}\label{impexp}

In \citet{Ju2009}, the strength and fracture model by \citet{BA1994,BA1995} in combination with a porosity model \citep{Ju2008} was compared to laboratory impact experiments involving porous pumice. Here, we present a similar comparison test of our updated strength model using the results of a study by \citet{Ok2009}, who conducted impact experiments using porous gypsum spheres and a wide range of specific energies (from 3$\times$10$^3$ J$/$kg to 5$\times$10$^4$ J$/$kg) and investigated the resulting fragment mass distributions. We use run 1 and 2 of this study with the initial conditions indicated in Table 1. 

As far as possible, we use material parameters which were measured for the target material.  For the crush-curve, we use a quadratic form (see \citealt{Ju2008}) with $P_e$ = 1$\times$10$^8$ dyn/cm$^2$ and $P_s$ = 4$\times$10$^9$ dyn/cm$^2$, which gives a reasonable fit of the measured crush-curve of the gypsum material used in the experiments (Okamoto and Arakawa, private communication). Further more, the initial slope of the crush-curve in the elastic regime is chosen to match the measured longitudinal speed of sound of $\sim$ 2 km/s \citep{Ok2009}. The Weibull parameters were varied in the simulations in order to obtain a match of the largest remnant for one experiment (run 1). For the best fit model\footnote{The values for $m$ and $k$ found in this study are different from the ones obtained by \citet{BA1994,BA1995}, because a different kind of material is considered here.}, $m$ = 9.5 and $k$ = 8$\times$$10^{37}$. For the other runs, the same values of $m$ and $k$ were used. Note that the minimum strain threshold for failure resulting from these parameters is about 3$\times$$10^{-5}$ which, multiplied by the Young modulus of 5.3$\times$$10^{11}$, leads to an equivalent tensile strength of $\sim$ 1.5 MPa. This value is slightly smaller, but of the same order as the measured tensile strength of $\sim$ 2.5 MPa \citep{Ok2009}. In all cases, we use a coefficient of friction (damaged material) $\mu_d$ = 0.55, a coefficient of internal friction (intact material) of $\mu_i$ = 1.5 and a limiting yield strength of $Y_m$ = 3.5 GP.

\begin{figure}
\begin{center}
\includegraphics[width=12cm]{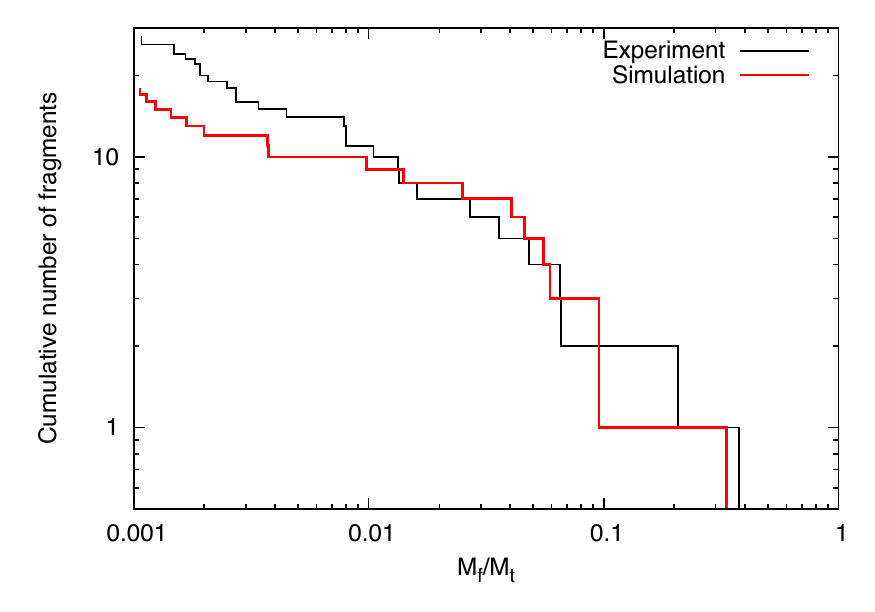}
\includegraphics[width=12cm]{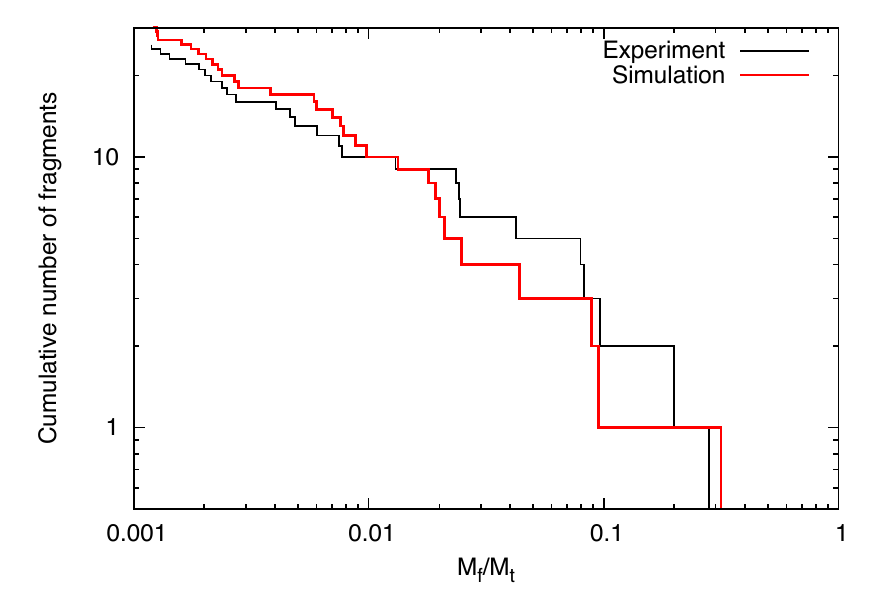}
\caption{Cumulative fragment size distributions obtained in the impact experiment by \citet{Ok2009} (black) and the SPH simulation (red). The targets consist of  porous gypsum; the impact conditions correspond to run 1 (top) and run 2 (bottom) of the experiments described in \citet{Ok2009}.}\label{fig:massdistr}
\end{center}

\end{figure}

\begin{figure}
\begin{center}
\includegraphics[width=12cm]{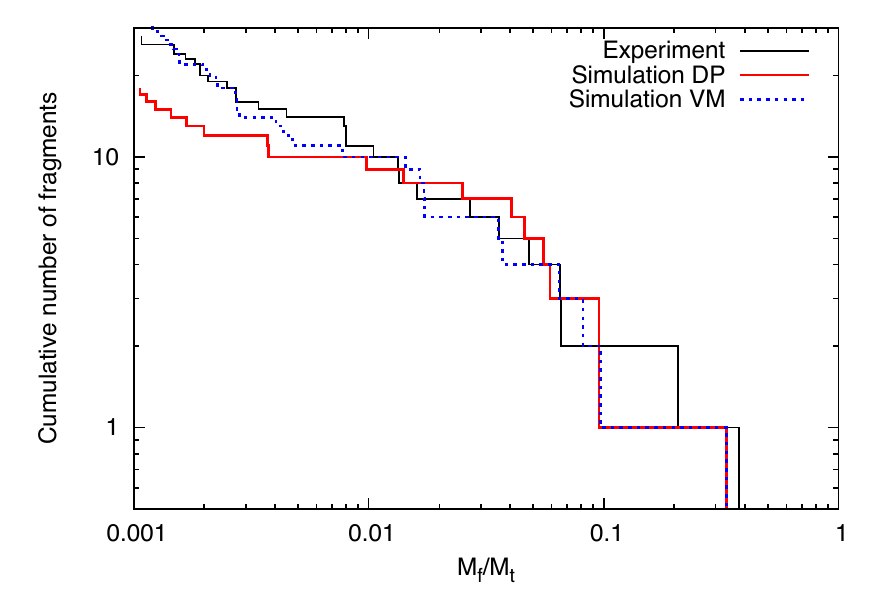}
\caption{Same as Figure \ref{fig:massdistr}, but only run 1 is shown. In the SPH simulations, the Drucker-Prager (DP) and von Mises (VM) strength models are compared.
}\label{fig:massdistr_variousp}
\end{center}

\end{figure}

The results of our simulations compared to the experiments are presented in Figure \ref{fig:massdistr}. The experimental mass distribution is well reproduced in both cases. In Figure \ref{fig:massdistr_variousp}, the results of two simulations of run 1 using different yield criterions are compared. Our results indicate that, as long as the same limiting yield strength $Y_m$ is used, the Drucker-Prager and the originally used von Mises yield criterion lead to very similar results.
Besides the fragment size distribution, also the antipodal velocities of the fragments was measured. In the simulations, we obtain fragment velocities at the antipodal side varying between 4-5 m/s, which is in good agreement with the experimental values of $\sim$ 5 m/s for run 1 and 2. At the center of the antipodal side, we obtain in the simulation also some fully damaged material which is ejected with slightly higher speeds ($\sim$ 10 m/s).

\section{Catastrophic disruptions}
The catastrophic disruption energy threshold $Q^*_D$ was studied in the past for targets with various properties and structures using different previous versions of SPH codes (see Section \ref{s:intro}). Here, we present results of a new study of $Q^*_D$ using the updated material models (see Section \ref{sec:improv}) and investigating the relative effects of friction, porosity and strength. Self-gravity is included throughout these simulations using a grid-based gravity solver. 
To identify the largest fragment formed by reaccumulation of smaller pieces, we use an iterative procedure based on energy balance (Benz and Asphaug 1999). While this method does not allow to compute the whole size distribution of fragments, it was shown to be accurate in determining the largest reaccumulated fragment\citep{Ju2010}, as long as it has a significant size ($M_{lr}/M_{tot} > $ 10 - 20\%).

\subsection{Comparison to previous results}
 As a first step, we want to reproduce the results by \citet{BA1999} for solid basalt targets using the updated strength model. We find that, when starting with a intact homogenous solid target, both the old (von Mises) and new (Drucker-Prager like) strength models lead to the same (within a few \%) results in terms of the catastrophic disruption threshold $Q^*_D$, as long as the same limiting yield strength $Y_M$ is used. Note, however, that this is true only in the case where we start with a \emph{intact} solid target (as it was done in the  \citealt{BA1999} study). 
\subsection{Investigation of various target properties}
Our newly implemented material models allow us to investigate targets with various properties. In particular, thanks to the friction model, we can now also study collisions among initially fully or partly damaged targets (e.g, granular or rubble pile bodies).  In order to systematically study the relative effects of friction, porosity and cohesion, we investigate the disruption threshold $Q^*_D$ as a function of radius for spherical bodies with the following properties:
\begin{enumerate}
\item no friction, no crushing, no cohesion (purely hydrodynamic)
\item friction included, no crushing, no cohesion 
\item friction and crushing included, no cohesion 
\item friction,  crushing and cohesion included (an intact porous body, similar to pumice)
\end{enumerate}
In all cases, we start with the same initial density $\rho$ = 1.3 g/cm$^3$. Targets labeled 'no crushing' also have the same initial density, but there is no crushing of pore space (i.e., the initial distention is set to $\alpha_0$ = 1). In the cases that include crushing, a crush-curve with parameters for pumice is used \citep{Ju2009} which means that the bodies are compacted (i.e., pore space is decreased) during the collision. In all the cases, we use a coefficient of friction (damaged material) $\mu_d$ = 0.8 and a limiting yield strength of $Y_m$ = 3.5 GPa. In the case with cohesion (target 4), a coefficient of internal friction (intact material) of $\mu_i$ = 1.5, and a cohesion $Y_0$ = 100 MPa are used. 

\begin{figure}
\begin{center}
\includegraphics[width=14cm]{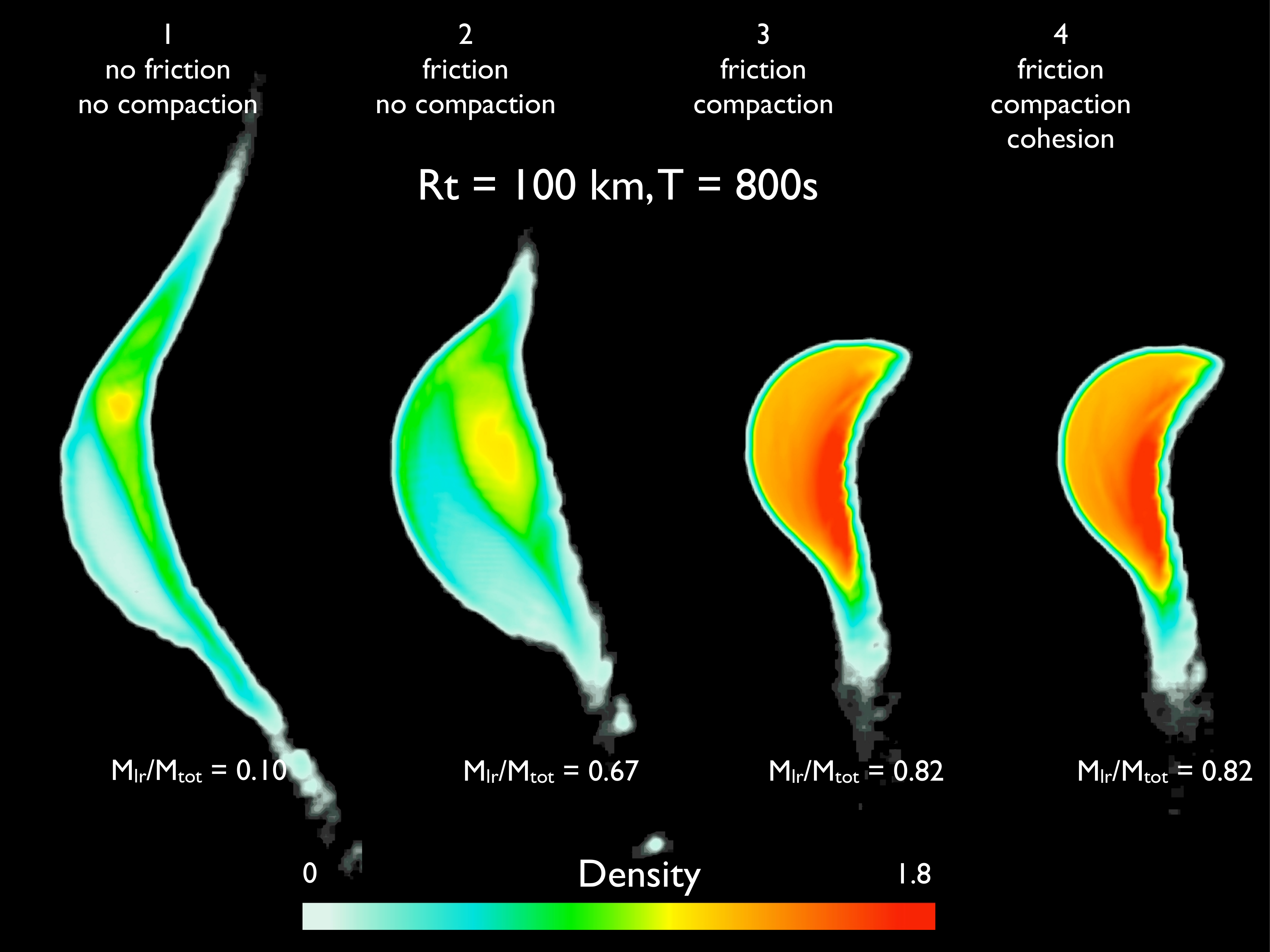}
\caption{Cross-section of SPH simulations of collisions between a target with a radius of $R_t$ = 100km and projectile of $R_p$ = 27 km with a relative velocity of 3 km/s and a 45$^\circ$ impact angle. Four different targets 1 - 4 (as indicated at the top) are investigated. The outcomes in terms of the degree of disruption, size of the largest remnant ($M_{lr}/M_{tot}$) and density increase strongly depend on the target properties.}\label{fig:R100kmComp}
\end{center}

\end{figure}

\begin{figure}
\begin{center}
\includegraphics[width=14cm]{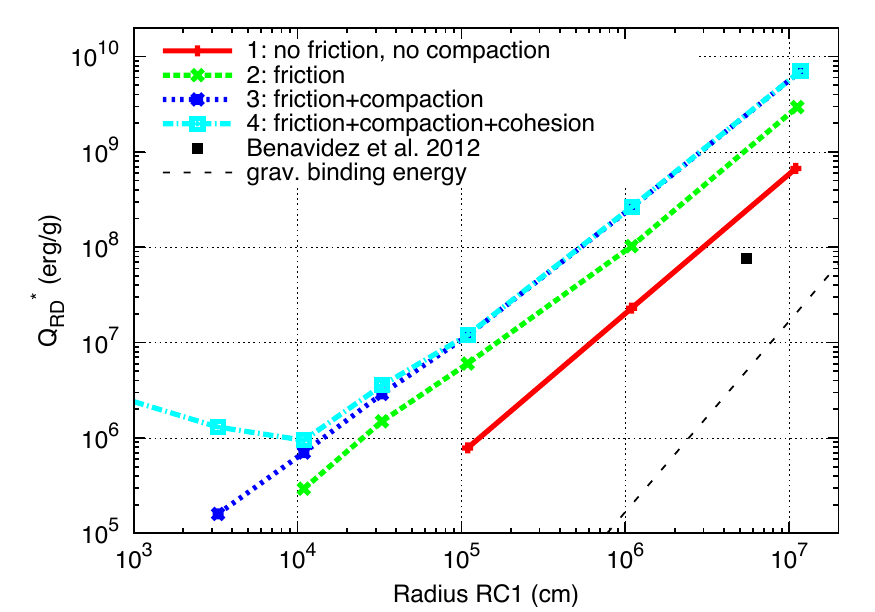}
\caption{Catastrophic disruption threshold $Q^*_{RD}$ as a function of the combined radius $R_{C1}$ for various target properties. The $Q^*_{RD}$ curves are the result of new SPH code calculations. The black point corresponds to the value of $Q^*_{RD}$ estimated from  simulations of "rubble-pile" collisions by \citet{Be2012}.}\label{fig:qcrit}
\end{center}

\end{figure}

For the simulations presented here, we use an impact velocity of 3 km/s and an impact angle of 45$^\circ$. In Figure \ref{fig:R100kmComp}, the outcome of a collision using a target with radius of $R_t$ = 100 km, and an impactor with $R_p$= 27 km is shown at the time $t$ = 800s for the cases 1-4. As it can be seen in the cross sections shown in Figure \ref{fig:R100kmComp}, the degree of disruption and consequently the mass of the largest remnant $M_{lr}$ after the collision strongly depend on the target properties. The 'hydrodynamic' target experiences the highest degree of disruption and the resulting largest remnant is the smallest ($M_{lr}/M_{tot}=0.10)$. In the case 2 where friction is included, the degree of disruption is much smaller and $M_{lr}/M_{tot}=0.67$. In the cases 3 (with friction and crushing) and 4 (with friction, crushing and cohesion), the target is significantly compacted by the collision and the density is increased accordingly. In both cases, $M_{lr}/M_{tot}=0.82$, which is larger than for the other targets. The outcome for runs 3 and 4 are very similar, which indicates that for disruptive collisions at this size scale, the effects of cohesion and tensile strength are negligible. On the other hand, porosity (energy dissipation by compaction) and friction (when starting with a damaged body) are very important. Figure \ref{fig:qcrit} shows the disruption threshold $Q^*_{RD}$  for the for targets 1 - 4 as a function of the combined target radius $R_{C1}$ (defined in section \ref{s:intro}). The $Q^*_{RD}$ curves confirm the results discussed above for the $R_t$ = 100 km case. The hydrodynamic target has a by far the lowest $Q^*_{RD}$. The disruption threshold is significantly (5-10 times) higher when friction is included and is further increased (by a factor of 2-3) when the energy dissipation by compaction (pore crushing) is taken into account. The curves for the targets 3 and 4 are almost exactly the same for target sizes $R_t \ge$ 1 km. For smaller targets, cohesion and tensile strength start to affect the outcome and at even smaller scales, the largest fragments in the case of the cohesive targets are intact fragments rather then purely gravitational aggregates, which explains the deviation of the curves of the targets 3 and 4 at small sizes (strength regime). As it was found in previous studies, the $Q^*_{RD}$ increases with decreasing size for cohesive targets in the strength regime due to the size dependence of the tensile strength. 
In the Figure \ref{fig:qcrit}, we also plot one point representing the $Q^*_{RD}$ for the $R_t$ = 50 km rubble pile bodies used in the study by \citet{Be2012}. In this study, the rubble pile targets were constructed by filling the interior of a 100-km-diameter spherical shell with an uneven distribution of solid basalt spheres having diameters between 8 km and 20 km. It is important to note that this study was performed using an old version of the SPH code which does not include friction and which uses the von Mises yield criterion. As shown in Figure \ref{fig:qcrit}, the rubble pile targets used in the \citet{Be2012} study have a very low $Q^*_{RD}$ which even falls below the curve of our hydrodynamic targets. 
\section{Discussion and outlook}
In this paper, we presented recent improvements of the SPH technique concerning the modeling of the disruption of strength dominated bodies. The updated models are able to reproduce the results of laboratory experiments very well. As for the modeling of catastrophic disruptions of asteroids, a comparison between old and new strength models shows no significant deviation in the case of targets which are initially non-porous, fully intact and have a homogenous structure (such as the targets used in the \citealt{BA1999} study). In this case, the crucial parameter is the limiting yield strength $Y_M$ (see also \citealt{LS2009}), and the details of the strength model (e.g., pressure dependent vs. pressure independent yield strength) do not play an important role. At large scales ($>$ 1 km) the bodies get fully damaged in the disruptive collisions investigate here. It was found that in this case, the effect of friction in the post-impact flow (after the shock wave has damaged the material) is negligible. However, it is important to point out that this is not true in many other cases (e.g., initially partly or fully damaged targets, rubble-pile structures, etc.) in which it is crucial that the damaged material still has a pressure dependent shear strength (i.e, there is friction). As our investigations of the disruption threshold for different target properties show, an initially (pre-impact) fully damaged target modeled without friction or compaction has a $Q^*_{RD}$ which is significantly (5-10 times) smaller than the $Q^*_{RD}$ for the same target but where friction is taken into account. Interestingly, the rubble-pile targets used in the study by \citet{Be2012} (who used an older version of the SPH code) seem to have a $Q^*_D$ which is very low and even lies below our $Q^*_D$ curve for the purely hydrodynamic bodies. Our results therefore indicate that these bodies represent something like a porous fluid rather than real rubble-pile bodies, which we believe should have a $Q^*_D$ lying between the curves of the cases 2 and 3 in Figure \ref{fig:qcrit}. 
As our investigations show, when the  effect of the energy dissipation due to compaction (pore crushing) is taken into account, the targets become stronger ($Q^*_{RD}$ is increased by a factor of 2-3)\footnote{Note that \citet{Ju2010}, in a comparison between porous pumice and solid basalt targets, found that at large scales (gravity regime), the non-porous targets have a higher $Q^*_{RD}$. This different behavior is due to differences in density and shear strength ($Y_M$)}. On the other hand, cohesion is only important at small scales ($\lesssim$ 1km) and has an negligible effect at larger scales.
Our results therefore indicate that in the gravity regime, the crucial parameters are the crushing properties (crush-curve), 
friction (when starting with partly or fully damaged bodies) and the limiting yield strength $Y_M$ (the effect of varying $Y_M$ was already investigated by \citealt{LS2009}). Cohesion and tensile strength do not play an important role in this regime. 

Our results clearly confirm that collisions between bodies of a few 100 km diameter can not be treated by just using 'hydrodynamic objects', even if they are fully damaged without cohesion. However, at a certain size, bodies are expected to be fully gravity dominated (due to the increase of overburden pressure with size); the transition to this regime will be subject of a future study. 

The results presented here, as well as those from an ongoing, more general study (where a much larger parameter space will be covered) will be used to improve existing scaling laws for the outcome of collisions (e.g. \citealt{LS2012}).

\section*{Acknowledgments}
We thank C. Okamoto and M. Arakawa for providing the data of the gypsum impact experiments, W. Benz and E. Asphaug for developing the original SPH models and P. Michel for fruitful discussions. We also thank two anonymous reviewers for valuable comments.
M.J. is supported by the Ambizione program of the Swiss National Science Foundation.

\clearpage
\clearpage

\clearpage
\clearpage

\clearpage
\clearpage

\clearpage
\clearpage

\clearpage
\clearpage

\begin{table}[ht!]
\caption{Initial conditions and results of impact experiments \citep{Ok2009}. $m_l/M_t$ is the largest fragment mass normalized by the original target mass ($M_t$). $V_i$, $Q$ and $V_a$ are the impact velocity, the specific energy and the antipodal velocity, respectively.}
\vspace{.25 truecm}
\begin{tabular}{llllllll}
Run number & $M_t$ (g)& Target diameter (mm)& $m_l/M_t$&$V_i (km/s) $&$Q$ (J/kg)&$V_a (m/s)$\\
\hline
No. 1 & 12.1 & 28.8 & 0.377 & 3.33  & $3.21\times10^3$ & 5.08 \\
No. 2 & 8.44 & 25.5 & 0.281 & 3.30  & $4.51\times10^3$ & 5.00 \\
\hline
\end{tabular}
\label{initialcond}
\end{table}

\clearpage


\end{document}